# Assessing Olfaction Using Ultrasonic Vocalization Recordings in Mouse Pups with a Sono-olfactometer

Sébastien Wagner [#], Pierre-Marie Lledo [#, *] and Françoise Lazarini [#, *]

Institut Pasteur, Perception and Memory Unit, UMR 3571, CNRS, Paris, F-75015, France
*For all correspondences: pmlledo@pasteur.fr; lazarini@pasteur.fr
[#]Contributed equally to this work

**[Abstract]** Olfaction is the first sensory modality to develop during fetal life in mammals, and plays a key role in the various behaviors of neonates such as feeding and social interaction. Odorant cues (*i.e.*, mother or predator scents) can trigger potentiation or inhibition of ultrasonic vocalizations (USV) emitted by pups following their isolation. Here, we report how USV are inhibited by olfactory cues using a sono-olfactometer that has been designed to quantify precisely olfaction in pups congenitally infected by cytomegalovirus. This olfactory-driven behavioral test assesses the USV emitted in presence of unfamiliar odorants such as citral scent or adult male mouse scent. We measure the number of USV emitted as an index of odorant detection during the three periods of the 5-min isolation time of the pup into the sono-olfactometer: first period without any odorant, second period with odorant exposure and last period with exhaust odorant. This protocol can be easily used to reveal olfactory deficits in pups with altered olfactory system due to toxic lesions or infectious diseases.

**[Background]** In mammals, olfaction is the first sensory sense to become functional *in utero*, long before audition and vision (Stickrod *et al.*, 1982; Sarnat and Yu, 2016). Survival and growth of neonates depend on the mother and rely heavily on their reciprocal olfactory-driven behaviors such as nipple localization, feeding, attachment, predator avoidance, *etc*. (Teicher and Blass, 1976; Brunjes and Alberts, 1979; Bell and Smotherman, 1980; Brouette-Lahlou *et al.*, 1992; Shair *et al.*, 1999; Hongo *et al.*, 2000; Perry *et al.*, 2016; Al Aïn *et al.*, 2017). Thus, congenital or perinatal impairment of the sense of smell, such as toxic or infectious injury of the olfactory system, could have profound health concerns. While a great number of non-operant and operant behavioral tests are available for assessing the sense of smell in adult rodents (Bodyak and Slotnick, 1989; Slotnick and Restrepo, 2005; Kobayakawa *et al.*, 2007; Yang and Crawley, 2009), there is a serious limitation in exploring olfaction in very young rodents due to their limited behavioral repertoire. Nevertheless, a suitable test to address odor perception in rodent pups has been designed based on the recording of ultrasonic isolation calls (Hofer and Shair, 1991; Hofer *et al.*, 2002; Lemasson *et al.*, 2005; Lazarini *et al.*, 2018). Young pups, while isolated from the mother and littermates and placed in low ambient temperature produce ultrasonic vocalizations (USV) at a high rate (Smith and Sales, 1980; Branchi *et al.*, 1998; Castellucci *et al.*, 2018), that promote maternal behavior such as searching for pups, retrieving and licking of pups (Noirot, 1974; Brounette-Lahlou *et al.*, 1992; Brunelli *et al.*, 1994). Infants of most mammals, including humans, also emit repeated vocalizations in the audible range after isolation, as a distress signal aiming at eliciting maternal behavior. USV emission of the isolated rodent pups stops at the contact of the mother, littermates or nest odor (Szentgyörgyi *et al.*, 2008). On the one hand, potentiation of the USV response to isolation can be induced by exposition to the scent of its mother or another lactating female (Shair *et al.*, 1999). On the other hand, inhibition of the ultrasonic calls could be induced by exposition to the scent of an unfamiliar adult male, one of its common predators in the wild (Shair *et al.*, 1999). USV inhibition can also be achieved by exposure to a non-social odorant cue, such as citral, that triggers innate aversive response (Lemasson *et al.*, 2005). Using this olfactory-induced USV inhibition, we found that congenital cytomegalovirus (CMV) infection alters olfaction as early as Day 6 after birth, long before hearing deterioration in mice (Lazarini *et al.*, 2018).

The current protocol describes a method for assessing odor perception in young mouse pups soon after birth, using a custom-made sono-olfactometer. Because this approach has been developed in the context of congenital viral infection, the sono-olfactometer was designed to prevent any spread of the virus from infected pups to the environment. It can be used therefore in Biosafety level 1 (BSL-1), level 2 (BSL-2) or level 3 (BSL-3) according to the microbial status of the manipulated animals. This characteristic was made possible by slightly modifying the earlier version of our olfactometers (Lemasson *et al.*, 2005). The pup chamber of the sono-olfactometer constitutes a mini-isolator in which odorants can be presented at a constant concentration and then efficiently exhausted. This sono-olfactometer allows simultaneous exposition to various odorants and recordings of USV emitted by the pup placed in the chamber. This protocol can be easily expanded to explore olfaction in other paradigms of acute and chronic injury or infectious diseases in the olfactory system of wild-type or genetically-modified rodents.

**Materials and Reagents**

1. 50-ml tube (Corning, France, catalog number: 430828) with two custom-made 5 mm-diameter holes in the lid
2. Laboratory-bred mouse pups from 6-8 days after birth
   *Notes:*
   *a. Put in a single cage each pregnant female one week before the timed day of birth.*
   *b. For the test, male and female pups can be used. We only tested Oncins France 1 (OF1) mouse line from Charles Rivers, France with this behavioral protocol. This mouse strain is productive and widely used for teratology. Pups of other mouse strains such as C57Bl/6J emit similar USV* (Castellucci *et al.*, 2018).
   *c. Infected-pups can be used as previously described (Lazarini et al., 2018). OF1 mother mice and its litter from Charles Rivers, France were individually housed in two isolators, one for the control (CTL) group and the second for the CMV group, kept in a BSL-2 room with controlled temperature (22 °C) and humidity (range: 40%-70%), under 12 h light/dark cycle*

*(lights on at 8:00 AM) in the Pasteur Institute animal facilities accredited by the French Ministry of Agriculture for performing experiments on live rodents. Mice were manipulated in class II safety cabinets.*

  d. *You can identify the pups at Day 1 after birth using long-lasting paw tattoos, subcutaneously injected with a 0.3 mm x 13 mm needle.*

3. Male scent (10 g soiled bedding from a group of 6 unfamiliar male adult OF1 mice)
4. Mineral oil (Sigma, France catalog number: M5904)
5. Citral (Sigma, France catalog number: W230316)

   *Note: Citral has lemon scent.*

6. 70% Ethanol solution
7. Citral solution (see Recipes)

**<u>Equipment</u>**

1. Class II safety cabinets
2. Custom-made sono-olfactometer (depicted in Figure 1)
   a. The audio recording system is composed of:
      i. A sound card recorder (PreSonus AudioBox iTwo) (Figure 1). The sound card can be replaced by any other commercial model. The recording potentiometer is adjusted to maximize the signal-to-noise ratio and avoid overloading.
      ii. A heterodyne bat detector (whose microphone has been moved into the chamber isolator using a BNC cable) (Magenta Bat5 Digital Bat Detector, RSPB, UK) (Figures 1 and 2). The heterodyne bat detector is set to the center frequency of the mouse vocalization: The volume control of the bat detector is adjusted in the middle to avoid background noise.

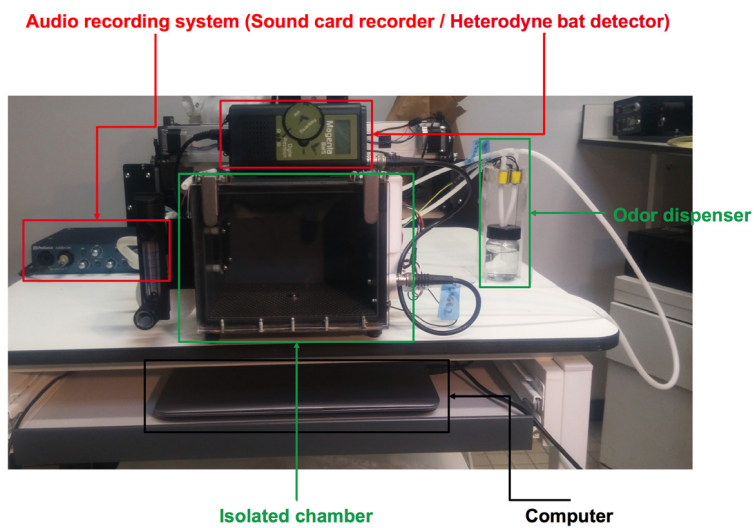

**Figure 1. The sono-olfactometer.** The sono-olfactometer is a system that delivers odors

while recording the ultrasound emitted by mice. It is composed of four sub-systems: 1) An audio recording system; 2) Two identical isolated chambers (only one chamber will be presented); 3) An odor dispenser (olfactometer); 4) A computer with its own software.

b. The isolated chamber (Figure 2):
   i. Exhaust air pump (Schego, catalog number: 850)
   ii. Non-return (check) valve (composed of two elements from Colder Products, catalog numbers: PLCD220-04 and PLCD10004)
   iii. Exhaust air flow meter (Brooks Instrument, catalog number: FR2A13BVBN)

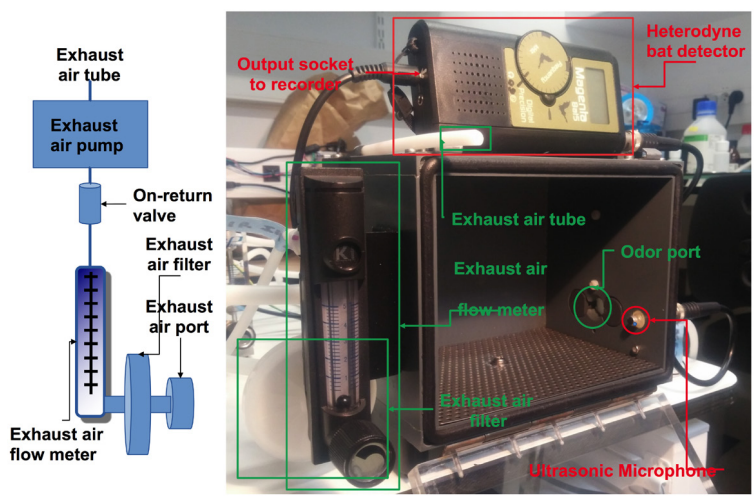

**Figure 2. The isolated chamber.** The isolated chamber (inside size 18 x 12 x 12 cm) is sealed and airtight when closed. The isolated chamber was made with components (8 mm-thick black PVC walls for the ceiling, the floor and sides, 8 mm-thick transparent plexiglass for the door, 5 mm-diameter rubber seal for the door, 0.1 mm-thick aluminum coating plate with 2 mm diameter holes) easily available in DIY stores such as Lacrylic shop, Bonneuil sur Marne, France. The odor port and the ultrasonic microphone are on the right side of the chamber. The odor dispenser releases the Odor/Air mixture in a controlled manner through the odor port (Figure 3). A drain hole in the center of the left side is connected to a HEPA exhaust air filter (Millex-FG, 0.20 µm, PTFE hydrophobe, 50 mm). The air is evacuated by an exhaust air pump, a non-return (check) valve (composed of two elements from Colder Products, and an exhaust air flow meter (Figure 3), thus avoiding microbial contamination of the environment. The exhaust air containing odorants is directly diverted to the air exhaust of the animal facility. The HEPA exhaust air filter is changed at the end of the experiment, after the testing of all the pups. The ultrasonic microphone is offset inside the box.

c. The odor dispenser (olfactometer) (Figure 3):
   i. Emitting air pump (Schego, catalog number: M2K3)

ii. Air flow meter (Key Instrument, catalog number: FR2A14BVBN)

iii. Odor flow meter (Brooks Instrument, catalog number: FR2A13BVBN)

iv. Odor valves (*e.g.*, ASCO, catalog number: SCH284A005.12/DC or Bio-Chem Fluidics, catalog number: 100P2NC12-05B or equivalent normally closed solenoid pinch valves)

v. Exhaust air pump (Schego, catalog number: 850)

vi. C-Flex® Standard Tubing (ID: 0.125 OD: 0.250)

*Note: The tubing is GMP Compliant.*

*Note: All electronics used to control the system is a custom-made device but it could easily be replaced by a commercialized version of Arduino card: https://www.arduino.cc/en/Main/Products.*

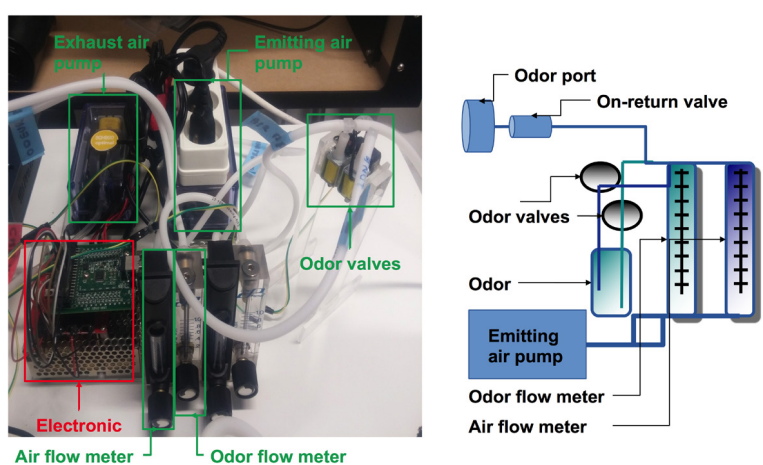

**Figure 3. The odor dispenser.** The odor dispenser distributes the Odor/Air mixture in a controlled manner. The air from the emitting air pump is transferred to the air flow meter and the odor flow meter. This device allows one to adjust the Odor/Air ratio (0.3 L/min for odorant and 2 L/min for air). If odor valves (can be "ASCO", "Bio-Chem Fluidics" or equivalent normally closed solenoid pinch valves) are closed, then clean air will be diffused into the chamber, otherwise the odor will be diffused homogeneously in the chamber. The air is evacuated into the isolated chamber by the exhaust air pump at a flow rate of 1.5 L/min. C-Flex® Standard Tubing (ID: 0.125 OD: 0.250) is used to connect the different devices.

3. Personal protective equipment (PPE)

   May include (but is not limited to) scrubs, a sterile combination, latex gloves, bouffant cap, ventilation mask, protective glasses, and shoe covers, depending on the regulation of the animal facility in which the work is taking place.

**Software**

1. The software (Figure 4):

   Pups emitted ultrasonic vocalizations at 40-120 kHz that were detected using an ultrasonic microphone connected to a bat detector (frequency range 10-130 kHz) that converts ultrasonic sounds to the audible frequency range. Using the broadband 60 kHz output of the detector, ultrasonic calls were sampled, recorded and analyzed using Audacity open software (www.audacityteam.org).

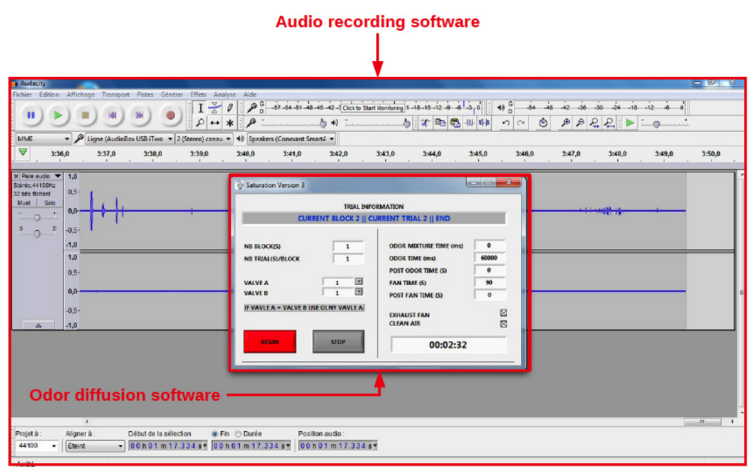

**Figure 4. The audio recording software.** This figure shows the computer screen with the open images of the audio recording and the odor diffusion software. The control of the diffusion of odors is ensured by a custom-made software. This software can be replaced by the Node-RED software coupled with an Arduino. As the Arduino appears as a Serial device, the Serial in/out nodes can be used to communicate with it (https://nodered.org/docs/hardware/arduino).

2. GraphPad Prism software (GraphPad Software, USA) is used for data analysis

**Procedure**

*Notes:*
1. *Gloves should be worn for all steps that involve handling mice, odorants, the sono-olfactometer and its chambers.*
2. *Prior to bringing the pups into the sono-olfactometer, prepare the odorants, the chambers (ensure cleanliness, connectivity of wires and odor dispensers, the bat detector and the computer software.*
3. *Prepare a sheet with all the animal information including animal numbers, feet marking, genotype, treatment and/or inoculum, chamber number, weight, etc.*
4. *Five minutes prior to testing, pups are moved from the homeroom and eventually from colony isolator to the class II safety cabinets in the testing room, in their home cages with their dams and litter. Two pups were placed into the two detachable chambers of the sono-olfactometer under the class II safety cabinets (one pup per chamber); the chambers with the pups were then put into the laboratory-made sono-olfactometer placed in the BSL-2 room, at proximity to the class II safety cabinets, for USV-recordings.*

5. *Each test session in the sono-olfactometer lasts 5 min. Pups can be tested two by two, on Days 6 and 8 after birth.*
6. *There is a maximum of one session per day, with each session comprised of exposure to only one scent.*
7. *Pups should be weighed after the test every day (balance placed in the safety cabinet) prior to be placed back to the nest. They should be separated from their mother less than 30 min in total (time including transfer to the chamber, testing, weighing and transfer to the home cage).*

**Recording of USV** (Figure 5)

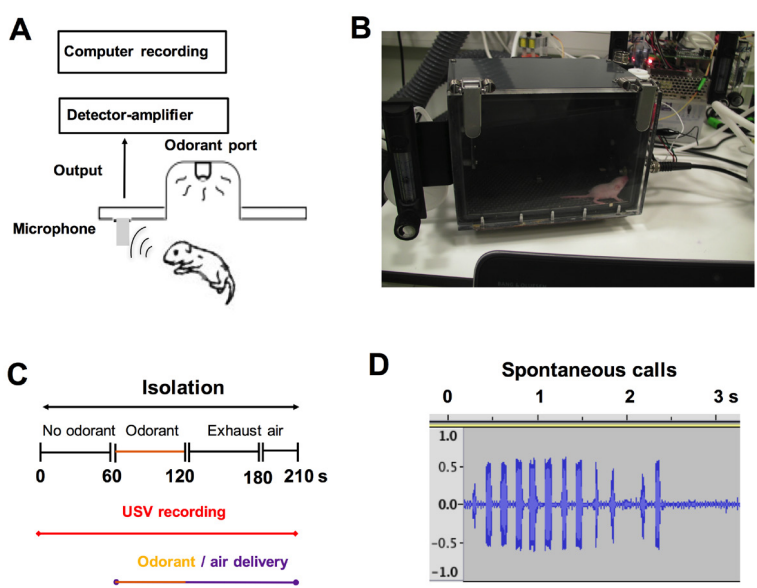

**Figure 5. Recordings and quantification of the emission of ultrasonic vocalizations.** A and B. The recording of ultrasonic calls began 30 s after placing the pups in the test chamber of the sono-olfactometer. Ultrasonic vocalizations were detected using an ultrasonic microphone connected to a bat detector that converts ultrasonic sounds into the audible frequency range (from 20 to 20,000 Hz). C. Experimental paradigm. Ultrasonic emissions were recorded during the first period without odorant (1 min), followed by a period of odorant exposure (1 min) and finally the last period of exhaust air (1 min and 30 s. This time duration allows the complete elimination of the exhaust air containing odorants). D. Typical wave traces of spontaneous call series from a pre-weaning 6-day-old pup (for more details, see Lazarini *et al.*, 2018). The majority of vocalizations (vocal units of duration < 100 ms on spectrogram) are produced in series with call intervals > 130 ms.

1. Place the two chambers containing the pups onto the table of the sono-olfactometer. Connect the sono-olfactometer (SO) and the computer. Switch on the bat detector.
2. Thirty seconds after connecting the two chambers to the sono-olfactometer, record the USV by starting the audio recording software. Record simultaneously the USV emitted by the two pups, each in their own sealed chamber. The routine protocol is shown in Figure 6. An example of USV recording is Sound 1 (this audio file depicts the USV of two 2-day-old pups).

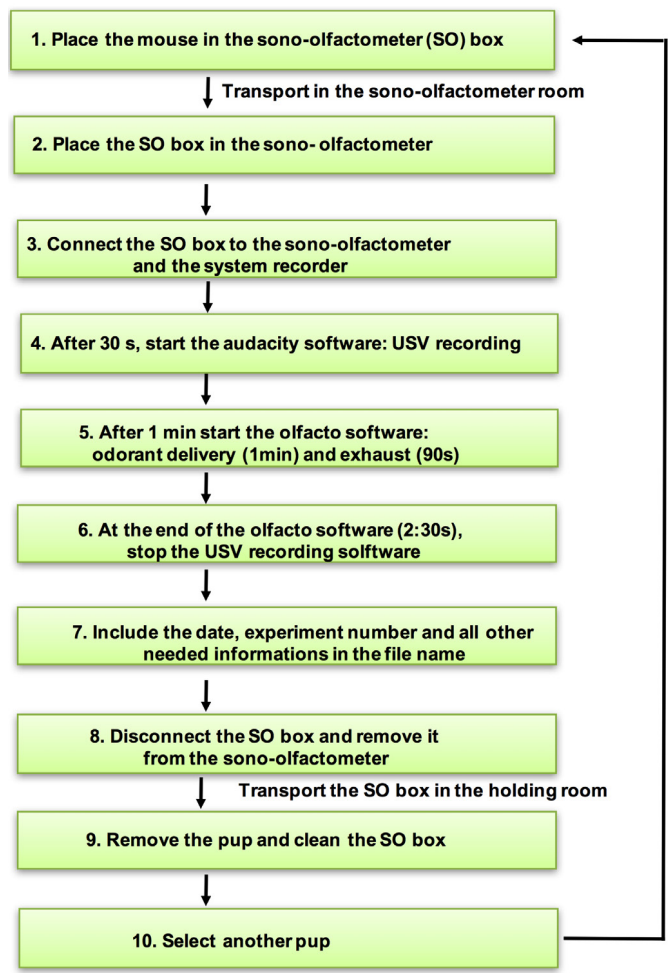

**Figure 6. USV recording in a sono-olfactometer.** The successive stages of the experimental procedure. Using the broadband 60 kHz output of the detector, ultrasonic calls were sampled, recorded and analyzed using the Audacity open software.

3. Sixty seconds after the beginning of USV recording, start the odor diffusion software.
4. When the odor diffusion program is finished (duration: 5 min), include the date, experiment number and all other needed information in the file name.
5. Transport the pups to their home cage (transfer from the chambers to the safety cabinet, weight them and transfer to the nest of the home cage).
6. Clean the chambers using 70% ethanol. After a 5-min interval to allow the elimination of the ethanol odorant, start the next test using the same chambers with other pups.
7. The USV records can now be analyzed using the Audacity software. An example of analysis is displayed in Figure 7.

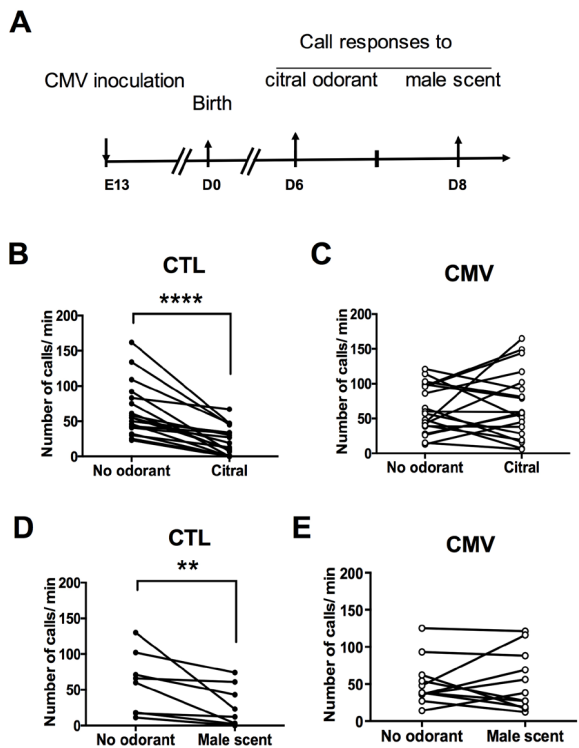

**Figure 7. Sample data of the olfactory USV inhibition test.** A. Timetable of the experiments. All pups in the same litter of timed pregnant mice were individually infected *in utero* at embryonic day 13 (E13) with intraplacental inoculation of murine CMV (Smith strain) under anesthesia. As a control (CTL) group, all pups in the same litter of other timed pregnant mice were intraplacentally injected with PBS at E13 under anesthesia. Animals were analyzed using sono-olfactometers on Days 6 and 8 after birth. B and C. Emission of ultrasonic calls for citral odorant on Day 6 after birth (n = 18 CTL, n = 19 CMV). D and E. Emission of ultrasonic calls for male scent odorant on Day 8 after birth (n = 8 CTL, n = 11 CMV). *P* values are calculated by Wilcoxon matched-pairs signed rank test. **$P < 0.01$, ****$P < 0.0001$; mean ± SEM in B-E. CTL pups decrease their emission of calls in response to exposure to non-social or social odorant molecules, such as citral or male scent, respectively. In contrast, congenital CMV infection impairs the ultrasonic call responses triggered by the two scents, indicating an alteration of olfactory perception induced by the virus (for more details, see Lazarini *et al.*, 2018).

## Data analysis

The number of ultrasonic vocalizations emitted after isolation was manually counted using Audacity open software (www.audacityteam.org). The mean rate of ultrasonic emissions (call/min) was computed for each time block: The first period without any odorant (1 min), the second period with exposure to social or non-social odorant (1 min) and the last period with exhaust odorant (1 min and 30 s). Data were analyzed using GraphPad Prism, using Mann-Whitney or Wilcoxon matched-pairs signed rank tests as appropriate.

**Notes**

1. Prior to testing pups, we recommend validating the two sono-olfactometer chambers by recording successively in each of them the same CTL pup aged of 2 days.
2. Since this test is a non-operant one (recording of spontaneous USV emitted by pups exposed to different olfactory cues), at least 8 pups per group should be used to reach the statistical power necessary for the analysis.
3. When manipulating two groups of animals, one infected and the other non-infected, we recommend dedicating each of the two SO chambers to each group using an external labeling. This should help in preventing any contamination of CTL pups.
4. The dilutions of Citral odorant should be made just before testing. We recommend preparing the scent under a fume hood in a laboratory outside the animal facilities since it is important to prevent the odor diffusion in the animal facilities (odor habituation).
5. For a greater reproducibility, fresh male bedding should be used before any microbial transformation of the scents.

**Recipes**

1. Citral solution
   1 ml Citral
   *ad* 10 ml mineral oil dilution

**Acknowledgments**

The predecessor to the presented sono-olfactometer (*i.e.*, the murine USV inhibition task based on an eight-channel olfactometer as described in Slotnick and Bodyak, 1999), was co-developed by Drs Morgane Lemasson, Gilles Gheusi and Pierre-Marie Lledo (first published as Lemasson *et al.*, 2005). The funding for the development of the sono-olfactometers was provided in a grant of the Institut Pasteur, Paris (GPF 2015 Microbes & brain "INFECSMELL") awarded to FL.

**Competing interests**

These results are the subject-matter of a U.S. provisional patent application number 62/793941 filed on 18 January 2019 on which Françoise Lazarini, Sébastien Wagner and Pierre-Marie Lledo are cited as inventors.

**Ethics**

All animal procedures were performed in accordance to the French legislation and in compliance with the European Communities Council Directives (2010/63/UE, French Law 2013-118, February 6th, 2013), according to the regulations of Inserm and Pasteur Institute Animal Care Committees. The Animal Experimentation Ethics Committee (CETEA 89) of the Pasteur Institute has approved this study (2015-0028). Mice were housed in isolators and manipulated in class II safety cabinets in the Pasteur Institute animal facilities accredited by the French Ministry of Agriculture for performing experiments on live rodents.